\documentclass[11pt]{article}
\usepackage{array}
\usepackage{graphicx}
\usepackage{textpos}
\usepackage{geometry}   
\usepackage{dsfont}
\usepackage{amsthm}
\usepackage{amsmath}
\usepackage{amssymb}
\usepackage{esint}
\usepackage{mathrsfs}
\usepackage{graphicx}
\usepackage{amssymb}
\usepackage{epstopdf}
\usepackage{textpos} 

\numberwithin{figure}{section}

\title{Particle Physics and Condensed Matter: The Saga Continues\footnote{Invited presentation of concluding remarks at Nobel Symposium 156 on New Forms of Matter, Topological Insulators and Superconductors, June 13-15 2014, H\"ogberga G{\aa}rd, Stockholm.}}

\author{Frank Wilczek\\
\small\it Center for Theoretical Physics, MIT, Cambridge MA 02139 USA}

\begin{document}

\maketitle

\begin{textblock*}{5cm}(11cm,-8.2cm)
\fbox{\footnotesize MIT-CTP-4787}
\end{textblock*}

\begin{abstract}
Ideas from quantum field theory and topology have proved remarkably fertile in suggesting new phenomena in the quantum physics of condensed matter.  Here I'll supply some broad, unifying context, both conceptual and historical, for the abundance of results reported at the Nobel Symposium on ``New Forms of Matter, Topological Insulators and Superconductors".  Since they distill some most basic ideas in their simplest forms, these concluding remarks might also serve, for non-specialists, as an introduction.
\end{abstract}

\medskip

\bigskip

This has been an extraordinary conference.  We've seen talks combining frontier mathematics, deep aspects of quantum mechanics, ingenious experiments, and visionary engineering.  We've seen strong links that connect us all the way from sophisticated versions of K theory to prototype circuit elements for quantum computers.   Long ago Isaac Newton spoke of  Analysis and Synthesis, as follows
\begin{quote}
As in Mathematicks, so in Natural Philosophy, the Investigation of difficult
Things by the Method of Analysis ought ever to precede the Method of Com-
position ... By this way we may proceed from Compounds to Ingredients, and
from Motions to the Forces producing them; and in general, from Effects to their
Causes. And the Synthesis consists in assuming the Causes discovered, and established as Principles, and by them explaining the Phenomena proceeding from
them, and proving the Explanations.
\end{quote}
Here we've seen Synthesis at its creative best.   

In these concluding remarks I'll try to do two things.  First, I'll very briefly reflect on the continuing symbiosis between ideas in high energy physics and condensed matter physics \cite{particleCondensed}.   It's an impressive series of success stories, in which topological insulators form the latest major installment \cite{topologicalInsulators}.  (Note: For this part of the talk I reference a few broad reviews; in the other part, I will supply references for a few works that are directly relevant to the text.)

Second I'll try to distill, more specifically, some key ideas that underlie the new results reported at the Symposium.   Here the emphasis will be on their deep but relatively simple roots, common to high energy and condensed matter physics, rather than their exuberant blossoming in particular materials.

\section{Meta-Themes}

{\it A priori\/} one might not have expected that ideas developed to describe the  behavior of interactions among elementary particles, as revealed at scales of $10^{-13}$ cm. or less, through studies using stupendous accelerators and detectors, could be transplanted into the description of emergent phenomena involving many thousands of molecules and scales billions of times larger, which are studied in modest laboratories and are the basis of practical devices and products.   The other direction, from large to small, might seem, if anything, even less promising.  

But let's look at the record.   

\begin{itemize}

\item{\it Quasiparticles}: Shortly after introducing photons as the elementary particles of light, Einstein applied his quantization idea to the vibration of crystals, giving birth to the concept of phonons -- the first emergent ``particles'' of condensed matter.  Dirac's hole theory was inspired, I suspect, by the theory of electrons in solids.   Landau developed the idea that low-energy excitations in a wide variety of condensed matter systems can be modeled as systems of interacting (quasi-)particles systematically, and made impressive applications.   
\item{\it Spontaneous Symmetry Breaking}:  Elemental facts of condensed matter physics, such as the existence of crystal structure and of ferromagnetism, exemplify spontaneous symmetry breaking.  But interest in spontaneously breaking itself (as opposed to particular realizations) came with recognition that it carries rich implications, including:
\begin{itemize}
\item the existence of soft modes, such as phonons in solids and magnons in ferromagnets, whose properties are connected directly with the broken symmetry.  Most impressively, the major phenomena of superfluidity can be understood on this basis.   In a remarkable feat of imagination, Nambu realized that $\pi$ mesons display properties suggestive of broken symmetry soft modes, and suggested what the underlying symmetry might be.   His general insights were important, historically, in making the possibility of {\it hidden symmetry}, which today dominates high energy physics, plausible and usable; and his specific suggestions remain fundamental for pion physics.  
\item the existence of characteristic defects, such as dislocations in crystals and domain walls in ferromagnets.  This symmetry-based perspective becomes especially fruitful in the study of liquid crystals.   The concept that topology might be responsible for the existence and structure of stable structures goes back to the prehistory of particle physics, with the vortex atoms of Lord Kelvin and their elaboration in Peter Tait's knot theory.   Their idea was that atoms are topologically stable defects in the lumineferous ether.   That particular classically inspired ether is no longer with us, but both materials and (superficially) empty space are ethers in a generalized sense, and they support a variety of stable defects. 
\item the existence of sharp phase transitions, accompanying changes in symmetry.  Understanding the singular behavior which accompanies phase transitions came from bringing in, and sharpening, sophisticated ideas from quantum field theory (the renormalization group).   The revamped renormalization group fed back in to quantum field theory, leading to asymptotic freedom, our modern theory of the strong interaction, and to promising ideas about the unification of forces.   The idea that changes in symmetry take place through phase transitions, with the possibility of supercooling, is a central part of the inflationary universe scenario.   
\end{itemize}
\item{\it Superconductivity}: The Bardeen-Cooper-Schrieffer (BCS) theory of superconductivity brought a new level of sophistication into many body theory.   Green function and Feynman diagram techniques that were highly developed in quantum field theory proved to be the natural language for many aspects of advanced condensed matter physics.   Conversely, appreciation that one could view superconductivity as a manifestation of gauged spontaneously symmetry breaking revealed new possibilities for modeling particle physics.   The Higgs mechanism is, in a real sense, simply superconductivity viewed from within the material.      
\item{\it Gauge Structures}:  Local symmetry was discovered, in the first instance, as a property of the fundamental equations of electrodynamics (and of general relativity).  In the generalized form discovered by Yang and Mills, it has proved a reliable guide to constructing theories of elementary particle interactions.   It is remarkable that macroscopic matter, too, manifests those mathematical ideals:
\begin{itemize}
\item global (topological) aspects of gauge theory became evident in Dirac's theory of magnetic monopoles, and in the closely related physics of the Aharonov-Bohm effect.  They are at the root of the exact quantization of magnetic flux in superconductivity, of resistivity in the quantum Hall effect, and the relation between frequency and voltage in the AC Josephson effect.  
\item emergent gauge structures are central to the theory of quantized Hall states, where they are closely connected to the physics of fractional statistics (anyons).   Though this circle of ideas, so far, has been difficult to access experimentally, I am optimistic about its future.  The possible utility of anyons for quantum information processing is an inspiring prospect.   (Fractional spin quantum Hall states, once they are engineered, will give us mobile, electrically neutral anyons, for which the effects of exotic statistics will be more salient.)  
\end{itemize}
\item{\it Topology}: I remember, as a student, being both puzzled and impressed by the shell model of nuclei.  How could such a crude caricature of complex, poorly-understood physics give so many useful quantitative results, in agreement with experiment?  Eventually I realized that one key ingredient is discreteness, which ultimately derives from topology -- to wit, the compactness of symmetry groups.  The point is that in predicting the value of quantities that are intrinsically discrete, small mistakes are correctable.  If you know that the quantity $x$ must be an integer, and your approximate calculation gives you $x= 1.1$, then either the correct value is $x =1$, or your approximation is very poor indeed.   When predicting  the ordering of energy levels in terms of quantum numbers like spin, parity, baryon number, and charge, the shell model is in this favorable situation.   The power of topology, to extract discreteness within continuity, and thus simplicity from complexity, has encouraged the development of intellectual tools that can be useful in many branches of physics, but have especially come into their own in condensed matter, with its richness of examples and abundant possibilities for creation.
\end{itemize}  


It's an impressive list -- and there's much more!  

\subsection{Quantum Matter: Bodies and ``Vacuum''}

The vibrant two-way intellectual traffic between such vastly different empirical domains gives us something profound to understand, as well as to celebrate.  

Why is this traffic possible?   There are several reasons:
\begin{itemize}
\item{\it Common principles}: Symmetry, locality, and the general framework of quantum theory are principles common to both fields, and they take us quite far, especially in conjunction with our next item.
\item{\it Encapsulation (quantum censorship)}: In principle a macroscopic object contains a multiverse of microcosms: solids are made from vast numbers of quarks, gluons, electrons, and photons.   One might expect that different, coarser methods would be necessary to describe the behavior of the multitude.  And indeed that is often the case, as we learn in statistical mechanics.   But at low energy quantum mechanics is a powerful censor, for it hides many degrees of freedom, locking them within substantial energy gaps.   Thus at different energy scales it is appropriate to regard atomic nuclei, atoms, molecules -- or materials -- as elementary building-blocks, all identical, locked into their ground states.   
\item{\it Materiality of ``empty'' space}:  We have learned, in high energy physics, that we must also be prepared to read encapsulation the other way.   The ground state of the world, which is commonly referred to as ``empty space'' or ``vacuum'', is in reality a complex material.  It is rich in condensates ({\it e. g}. the Higgs field, leading to mass generation) and quantum fluctuations ({\it e. g}. vacuum polarization, leading to asymptotic freedom), which largely condition the behavior of the epiphenomena -- ``particles'', ``matter'' -- we get to work with, and are.
\end{itemize}

Because these reasons are profound, we can expect our traffic to continue to expand.

\subsection{Toward Topological Insulators: Boundaries}

A big difference between high energy and condensed matter physics is that the latter offers a more varied range of accessible phenomena.   One kind of experiment, wherein one smashes together particles and monitors other particles coming out, dominates high energy physics.   Of course the {\it interpretation\/} of these experiments can -- and does -- involve a wealth of intricate ideas and structures.  But the phenomena, in themselves, are limited and rather crude.   

For example, I can't think of any real sense in which high energy physics can explore boundaries or interfaces in a controlled way experimentally, though they are central to many speculative models (``brane worlds'').  In condensed matter, of course, boundaries are not only unavoidable, but tremendously fruitful -- think of capacitors, diodes and LEDs, transistors, Josephson junctions, quantum Hall effect edge currents, ... .

In the remainder of this talk I'll how simple but profound ideas about possible behavior at boundaries, which are natural within relativistic quantum field theory, give an excellent introduction to topological insulators, and suggest some generalizations.  


\section{Ideas and Phenomena}

\subsection{Peirels' Instability and Its Relatives}

Consider a one-dimensional chain of identical molecules with spacing $a$.  We want to model the situation where there is one conduction electron per molecule.  Let us adopt the drastic simplifying assumption that the molecules simply provide a weak periodic background potential and ignore interactions among the electrons, aside from their mutual influence through fermionic quantum statistics.    To begin, we will also assume that the potential is spin-independent and invariant under spatial reflection.

According to standard band theory we should expect to have a metal, since the electrons exhaust only half the available states.  Indeed, the carrying capacity of the lowest band, allowing for the spin degree of freedom, is twice its extent in (quasi-)momentum space, and so the actual density corresponds to half filling:
\begin{equation}
N/L ~=~ 1/a ~=~  \frac{1}{2} \cdot 2 \cdot  \frac{\frac{2\pi}{a}}{2\pi} ~=~ 2 \cdot \frac{2k_F}{2\pi}
\end{equation}


\begin{figure}[h!]
\begin{center}
\includegraphics[scale=0.5]{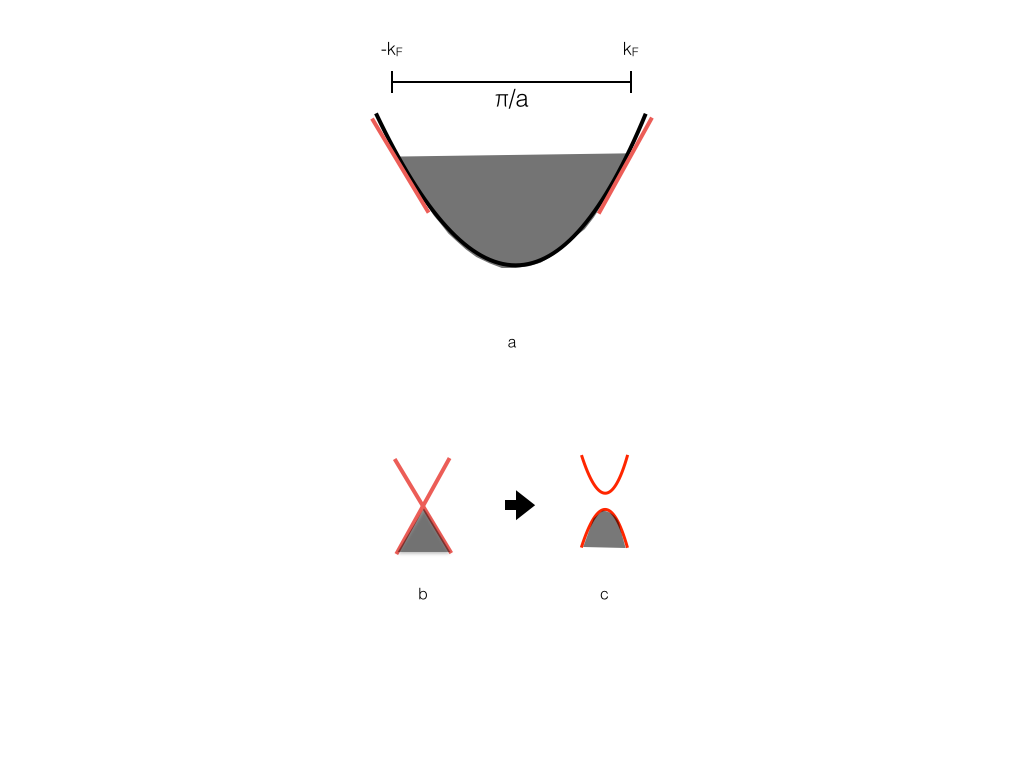}
\caption{a. Electrons at half filling in momentum space, for a one-dimensional lattice.  For concreteness, the drawing assumes that the lattice makes a weak perturbation to free electron behavior.   The Fermi sea is shaded.   Low energy excitations involve changes in occupancy near the Fermi points $k_F$, allowing us to linearize their dispersion relation.  b. The linearized spectrum can be mapped faithfully onto a light-cone.  c.  A deformation in the lattice structure which brings in momenta at $\pi/a$ can connect modes at $k_F + \xi$ and $-k_F + \xi$ -- right-moving particle and left-moving hole -- and open a gap in the dispersion relation.}
\label{peirelsMomentum}
\end{center}
\end{figure}

\bigskip

In the electrons' ground state, they occupy all the momentum states for $-k_F \leq k \leq k_F$ with
$k_F = \pi / 2a$.   (See Figure \ref{peirelsMomentum}a.)  The low-energy excitations around this ground state involve changes in the occupation of modes near the boundaries of that region.  We can approximate the change in their energy with momentum -- that is, the deviation in momentum from $\pm k_F$ -- as a linear relationship, bringing in the Fermi velocity
\begin{equation}
v_F ~=~ \frac{dE}{dk}(k_F)
\end{equation}
Thus we derive, from the two points of the Fermi surface, left- and right-moving excitations obeying
\begin{eqnarray}
(\partial_t - \partial_x ) \psi_R ~&=&~ 0 \nonumber \\
(\partial_t + \partial_x ) \psi_L ~&=&~ 0 
\end{eqnarray}
where we adopt the unit of velocity $v_F \rightarrow 1$.  This can be summarized in relativistic form, as a Dirac equation
\begin{equation}
i \gamma \cdot \partial \, \psi ~=~ 0 
\end{equation}
with
\begin{equation}
\psi ~\equiv~ \left(\begin{array}{c}\psi_L \\\psi_R\end{array}\right) \, ; \ \gamma_0 ~\equiv~ \sigma_2 \, : \ 
\gamma_1 ~\equiv~ i \sigma_1 
\end{equation}

This empowers us to visualize a Dirac cone, as displayed in Figure \ref{peirelsMomentum}b.  

The modes from the two sides of the cone are separated, as we saw earlier, by $\Delta k = \frac{\pi}{a}$.   This wavenumber is not represented in the background potential, so the modes remain independent.  

On the other hand, were the potential to contain that wave number it could connect the left- and right-movers, and open a gap in the Dirac cone, as displayed in Figure \ref{peirelsMomentum}c.  This mixing pushes down the energies of occupied states, so it is a favorable effect, energetically.    We should enquire whether it can be triggered dynamically, and occur spontaneously.   


\begin{figure}[h!]
\begin{center}
\includegraphics[scale=0.30]{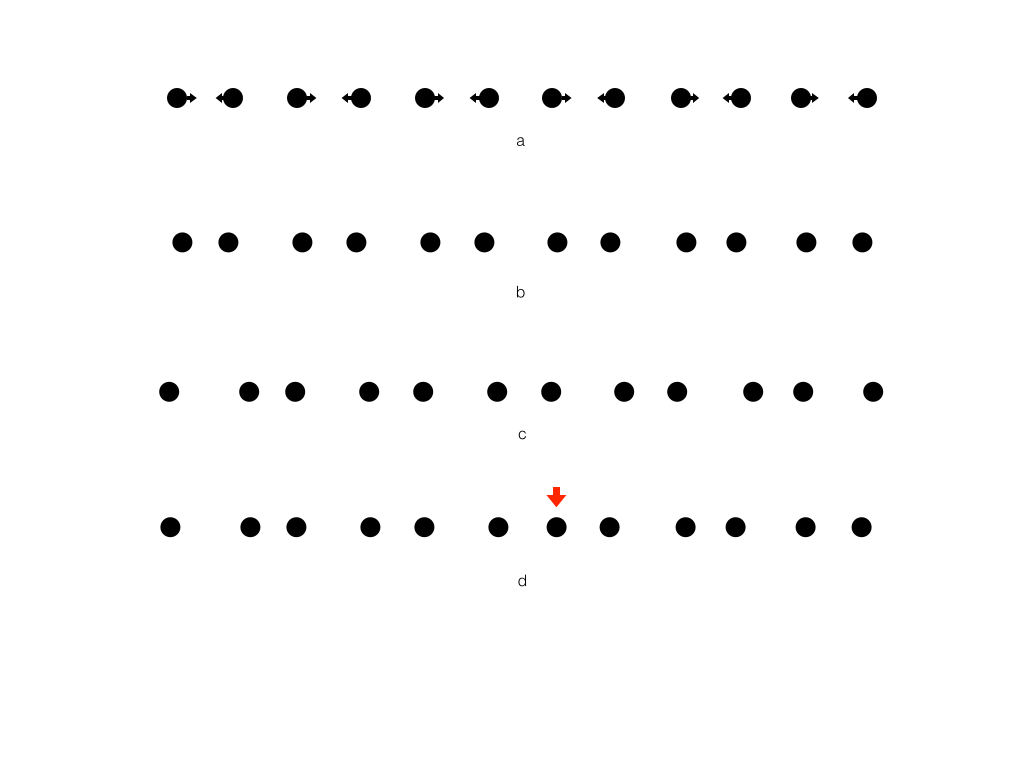}
\caption{a. Vibration of a periodic lattice at $\pi/a$, corresponding to an ``optical'' phonon.  b. and c.: Condensation in the optical phonon mode leads to a deformation of the lattice, doubling its periodicity to $2a$.  The condensation can occur in two different ways, as shown.   d. A simple slip in the deformation process, indicated by the arrow, produces a configuration which looks like the ordering of b. to the right and like c. to the left.  Such a configuration can relax into a domain wall, which is stable for topological reasons.}
\label{deformedLattice}
\end{center}
\end{figure}

\bigskip

To generate the required potential, we need to let the molecules displace in alternate directions, as displayed in Figure \ref{deformedLattice}a.    Displacements of the molecules correspond to phonons, and in that language we are asking whether the ``optical'' phonons at $k = \frac{\pi}{a}$ condense.  
If we denote the amplitude of the displacement (phonon)  $k = \frac{\pi}{a}$ field by $\phi (x, t)$, then we have a coupling
\begin{equation}\label{electronLagrangian}
{\cal L } ~=~ \bar \psi (i \gamma \cdot \partial - g \phi ) \psi 
\end{equation}
so that a condensation in $\phi$ opens a gap, as anticipated in Figure \ref{peirelsMomentum}c.  It gives us a fermion mass $m = g\langle \phi \rangle$.   

\begin{figure}[h!]
\begin{center}
\includegraphics[scale=0.5]{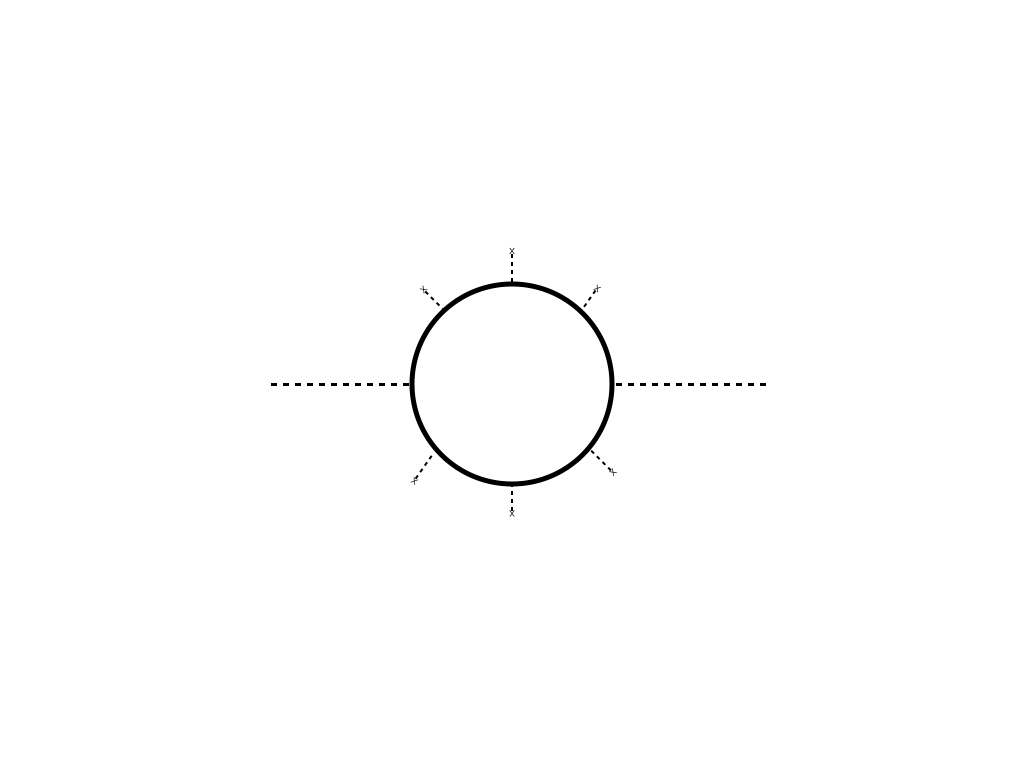}
\caption{This one-loop Feynman graph contributes to the effective potential for the optical phonon mode.  It captures the favorable energetics of opening a gap (or, in relativistic language, mass generation).}
\label{feynmanEnergy}
\end{center}
\end{figure}

This entire discussion can be phrased in the language of relativistic quantum field theory.  In that language, we are studying how mass generation back-reacts on the energetics of a scalar field which triggers it.   (In particle physics this effect is called the Coleman-Weinberg mechanism.  It plays an important role in the theory of the standard model, and in many speculations that go beyond it.)  We can exploit this mapping to bring in the machinery of Feynman graphs, easing our computational challenge.   Our energy gain corresponds to the vacuum amplitude shown in Figure \ref{feynmanEnergy}.    


\bigskip

To assess the possible instability, we can focus on infinitesimal displacements.  The most singular part takes the form
\begin{eqnarray}\label{displacementEnergy}
{\cal E} ~&\sim&~ \int\limits^\Lambda_0 \, \frac{d^2 k}{(2 \pi)^2} \frac{ {\rm Tr} ( \gamma \cdot k )^2}{(k^2 + m^2)^2} \nonumber \\ 
~&\propto&~  (g \phi)^2 \, \ln \, (\Lambda / g\phi ) 
\end{eqnarray}
Note that it is appropriate to supply an ultraviolet cutoff, since our simple ``relativistic'' description of the electron modes is only valid for low-energy excitations, involving small deviations from the Fermi points.   

Of course, displacement of the molecules also involves ordinary, Hooke's law, elastic energy.   But that energy, being proportional to $\phi^2$, is dominated by Eqn.\,(\ref{displacementEnergy}) for small $\phi$.   Thus the lattice will always deform: this is Peirels' instability \cite{peirels}.

\subsection{Domain Walls, and Fractional (Topological) Quantum Numbers}

The most favorable displacement amplitude $\langle \phi \rangle \equiv \pm v$ has two possible values.  They correspond to two homogeneous ground states.  They differ from one another by displacement through $a$ -- that is, through the transformation which used to be, but is no longer, a symmetry.  

In these two configurations our electrons have, according to Eqn.\,(\ref{electronLagrangian}), mass $\pm gv$.   This looks a bit strange, at first sight, but the sign of the mass can be absorbed, formally, into a redefinition of $\psi$.  Indeed, the transformation $\psi^\prime = \gamma_0 \gamma_1 \psi$ changes the sign of the mass term, while leaving the kinetic term invariant.   

A simple defect allows one to interpolate between the two distinct homogeneous states.  (See Figure \ref{deformedLattice}b.)  
If we let the simple defect configuration relax, energetically, it will settle down to a minimum energy configuration -- a domain wall -- that is stable for topological reasons.   

In this configuration the effective electron mass $m(x)$ interpolates between $\, -gv$ (say) on the far left and $gv$ on the far right.   (The {\it relative \/} sign, of course, cannot be undone by re-definition.)  The effective mass will vanish somewhere in between, and we might anticipate that something interesting might happen as a result.    In fact there is a remarkably robust and characteristic consequence: the existence of zero energy (midgap) states, centered on the wall.   

Fixing the center of the wall to be at $x=0$, we have an effective mass $m(x)$ with $m(-x) = - m(x)$.  The electron equation for zero energy reads
\begin{eqnarray}
0 ~&=&~ (- i\gamma_1 \partial_x  + m(x) )  \psi (x) \nonumber \\
~&=&~(\sigma_1 \partial_x  + m(x) )  \psi (x)
\end{eqnarray}
Projecting this on spinor structures with $\sigma_1 s_{\pm} = \pm s_{\pm}$, we find that for $\psi(x) \equiv \psi_+ (x) s_+ \, + \, \psi_- (x) s_- $
\begin{equation}
(\partial_x \pm m(x) ) \psi_\pm (x) ~=~ 0
\end{equation}
with solutions
\begin{eqnarray}
\psi^0_+ (x) ~&=&~ e^{- \int\limits_0^x \, du \ m(u) } \label{goodSolution} \\
\psi^0_-(x) ~&=&~ e^{ \int\limits_0^x \,  du \ m(u) }
\end{eqnarray}
The first of these, i.e. Eqn.\,(\ref{goodSolution}), defines a normalizable state, which shrinks exponentially to zero in both directions away from the wall.  It is the zero energy, midgap state advertised just above.  

The zero energy mode has several interesting consequences.  It should be plausible, given the symmetry of the problem, that the zero energy mode draws half its strength from below and gap half from above.    Thus if we fail to occupy the zero mode, we have a deficit of half a unit of electron charge, relative to the ground state; while if we do occupy the zero mode, we have an excess of half a unit of electric charge.   One can also show this formally, either directly from the definition of the charge operator, or by calculating the flow of charge as one builds up the domain wall (see below).  This domain wall induces a special kind of vacuum polarization, where a fractional charge accumulates \cite{jackiwRebbiSSH}.    

When we take into account the spin degree of freedom, the accounting takes a different form.  There are two zero modes: one for spin up, one for spin down.   By occupying, or not, each of the two zero modes, we have:
\begin{itemize}\label{funnySpinCharge}
\item {\it zero occupancy}: charge $-e$, spin 0
\item {\it single occupancy}: charge zero, spin $\frac{1}{2}$ doublet
\item {\it double occupancy}: charge $+e$, spin 0
\end{itemize}
Though the charge spectrum is normal, the relation between charge and spin is unusual.   We can express the general situation through the equation
\begin{equation}\label{chargeSpinTopologyRelation}
(-1)^{2S} (-1)^{Q/e} ~=~ (-1)^W
\end{equation}
where $W$ is a topological quantity, indicating the number of domain walls.   Since the quantities that appear in it are intrinsically discrete this relation will be exact, unless our approximations have been very bad.   Moreover, it is the only non-trivial relation among $2S, Q/e$, and $W$ consistent with the property that two domain walls can annihilate into states with normal quantum numbers.  We expect this property to be valid, because the lattice of molecules with two minimal defects differs from a correctly ordered (i.e., ground state) lattice only within a bounded region, and is topologically trivial.

The logic of the Peirels instability is not restricted to half filling.  One can, for example, consider conduction bands that fill $1/k$ of the available states, and trigger an instability toward charge density waves at $2\pi/k$, with $k$ an integer.   Then we will have domain walls that can annihilate in $k$-tuples.   If such annihilation is accompanied by emission or absorption of $l$ electrons, we can deduce a generalization of Eqn.\,(\ref{chargeSpinTopologyRelation}), in the form
\begin{equation}
e^{i2\pi (Q/e  - lW/k)} ~=~ 1
\end{equation}
(Here, for simplicity, I have not kept track of the spin quantum number, which need not be conserved separately.)  Equivalently, we can write
\begin{equation}\label{chargeWallRelation}
Q/e ~=~ l W  / k \, + \, {\rm integer}
\end{equation}
Earlier we had $k=2, l=1$, and half-integer charge.  In that case, we could understand the fractional charge based on the existence of a zero energy solution.  But in more general situations, say $k=3, l=1$, we can have third-integer charge, which cannot be understood in that way.

An appropriate, minimal model for these more general situations allows the field $\phi$ in Eqn.\,(\ref{electronLagrangian}) to become complex.   The overall phase of $\phi$ adjusted by re-defining $\psi$, according to $\psi^\prime = e^{i\lambda} \psi$, but relative phases in $\phi(x, t)$ are physically meaningful.   In particular, we can have situations where there are domain walls which interpolate between values
\begin{eqnarray}\label{wallAsymptotics}
\phi(x) \, &\rightarrow& \, \ \ v \, ;  \ \ \ \ \ \ \  x \rightarrow - \infty \nonumber \\
\phi(x) \, &\rightarrow&  \, e^{2\pi i/ k} \, v \, ;  \ \ x \rightarrow + \infty
\end{eqnarray}
Such domain walls can annihilate in $k$-tuples.   We expect, based on the preceding discussion, that fractional charge may accumulate on such walls.  

An efficient way to calculate the charge is first to imagine building up configurations with the wall asymptotics of Eqn.\,(\ref{wallAsymptotics}) gradually, starting from trivial asymptotics \cite{gw}.    As long as the magnitude of the local gap exceed the field gradients, i.e.
\begin{equation}\label{softWall}
\frac{ | \partial \phi | }{ g |\phi |} \, << 1
\end{equation} 
there will be no particle production, and we can calculate the current flow (to lowest order in gradients) by means of a simple vacuum polarization graph, similar in spirit to Figure \ref{feynmanEnergy}, but now with insertion of the electron number current in place of a phonon field. An elegantly simple result emerges, in the form 
\begin{equation}\label{gwCurrent}
\langle j^\mu \rangle \, = \, \frac{1}{2\pi} \, \epsilon^{\mu \nu} \, \partial_\nu \, {\rm Arg } \, \phi  
\end{equation}
and for the integrated charge
\begin{eqnarray}\label{gwRelation}
Q/e \, &=& \, \int\limits^\infty_{- \infty} dx j^0 \nonumber \\ 
\, &=& \,  \frac{1}{2\pi} \, \int\limits^\infty_{- \infty} dx \, \partial_x {\rm Arg} \, \phi \nonumber \\
\, &=& \, \frac{1}{2\pi} \, \bigl( {\rm Arg} \, \phi (\infty ) \,  - \, {\rm Arg } \, (- \infty) \bigr)
\end{eqnarray}

Now the realistic, minimum energy wall configuration may not satisfy Eqn.\,(\ref{softWall}) everywhere, though of course it does asymptotically (where $\partial \phi \rightarrow 0$).   So we must imagine a second step, where we build up the steeper gradients.  During that process electrons can be radiated to, or absorbed from, infinity.  But since the electrons are normal out there, any such radiation will change the electron number by an integer.   Also, the angle function ${\rm Arg} \phi$ becomes ill-defined at $\phi = 0$, and we need to allow for extra $2\pi$ jumps, as we integrate Eqn.\,(\ref{gwCurrent}) through such points.  For both these reasons, we should generally interpret Eqn.\,(\ref{gwRelation}) as a relation modulo integers, i.e. as a formula for the fractional part of the charge.  As such, it precisely embodies Eqn.\.(\ref{chargeWallRelation}).  

We can ``predict'' the existence of the zero energy solution we found earlier, based on these considerations, as follows.  The zero energy solution occurred in the model with $g$ and $\phi$ real.   Within that framework we cannot achieve the non-trivial domain wall asymptotics, where $\langle \phi \rangle$ change sign, without encountering a zero of $\phi$, where the requirement of Eqn.\,(\ref{softWall}) cannot be met.   We can get around that difficulty by adding a small infinitesimal imaginary piece to $\phi$, and then removing it at the end.   Since there is a gap, the limit is harmless.   But depending on the sign of the added piece, we will get $Q/e = \pm \frac{1}{2}$.   So there must be degenerate states with these quantum numbers, and therefore a zero energy mode of the electron field, whose occupancy (or not) distinguishes those charges.  

We can also contemplate incommensurate density waves, allowing domain walls that allow the change in ${\rm Arg}$, and thus the accumulated charge, to take on any value, rational or irrational. 

\subsection{Boundaries as Walls}

The possibility of zero-energy states at the termination of a 1-dimensional lattice was an early discovery of Shockley \cite{shockley}.  It gave a microscopically-based model of Bardeen's theory of ``surface states'', which in turn played an important role in the discovery of solid-state transistors.   Shockley's model is closely related to the models we have been discussing.  We can put his discovery into that framework, and generalize it, by considering how we might model boundaries of chains (or, of course, surfaces) as extreme versions of domain walls, as follows.   

Taking Eqn.\,(\ref{electronLagrangian}) as our basic model, we can have some value $g\phi_+$ (not necessarily real) for the effective mass in bulk, in for $x \, >  \, 0$, while taking $g\phi_- \, \rightarrow \, \infty$ for $x \, < \, 0$, to make it difficult for the electrons to penetrate there.   (If desired we can have a small bridging region, and let $\phi$ interpolate continuously between those values.)   Thus we realize the boundary as a specific kind of domain wall, to which our general analysis can be applied.   

Topological insulators, in their simplest form, fit neatly into this framework, as follows.   Assuming $T$ symmetry, we let $g$ and $\phi$ be real.   Then the relevant issue is the relative sign of $m(x)$ for between $x<0$ and $x>0$.  If the sign changes, we have a zero energy surface state; if not, not.    If we keep spin as a passive (bookkeeping) quantum number throughout, then we will get even numbers of zero energy states, as in Eqn.\,(\ref{funnySpinCharge}); but if there are significant spin-dependent forces, then spin is not a good label, and in general we will get an odd number.   

\subsection{Axion Electrodynamics and Fractional Landau Levels}

Now I will introduce another thread into the discussion, concerning a simple augmentation of electrodynamics.

\subsubsection{Axion Electrodynamics}

Axions are hypothetical particles which appear in theoretical proposals to explain the validity of $T$ symmetry in the strong interaction.  (General principles of relativistic quantum theory and local symmetry greatly constrain the possibilities for couplings among quarks and gluons, but they would allow a gluonic analogue of $\vec E \cdot \vec B$ to appear.  Nature does not exploit that option.  High energy physicists regard Her restraint as ``unnatural'', and look to new principles to explain it.   The most promising ideas lead one to predict the existence of axions.)   Axions are plausible candidates to provide the astronomical dark matter.  

Whether or not axions exist, they call attention to a set of equations that have interesting properties \cite{axionEM}.  One case of axion electrodynamics provides the effective theory for the interaction of topological insulators with the electromagnetic fields, and more general cases suggest natural generalizations.   There are also close connections to the physics of Landau levels and the quantum Hall effect, as we shall see.   

The central element of axion electrodynamics is the coupling (Lagrangian density)
\begin{equation}\label{axionEML}
{\cal L } ~=~ \kappa \, a \, \vec E \cdot \vec B ~=~ \frac{\kappa}{2} \, a \, \epsilon^{\alpha \beta \gamma \delta} F_{\alpha \beta} F_{\gamma \delta}
\end{equation}
between a (pseudo-)scalar field $a$ and the electromagnetic field.  For reasons that will appear below, it is useful to write
\begin{equation}
\kappa ~=~ \frac{e^2}{4\pi^2}
\end{equation}
To model particle physics axions one would include kinetic and mass terms for the axion field, but we will not do that here.  

If we add the coupling of Eqn.\,(\ref{axionEML}) to the standard Maxwell Lagrangian, then the resulting equations of motion read
\begin{eqnarray}
\nabla \cdot E ~&=&~ - \kappa \, \nabla a \cdot B \label{effectiveCharge} \\
\nabla \times E ~&=&~ -\frac{\partial B}{\partial t} \\
\nabla \cdot B ~&=&~ 0 \\
\nabla \times B ~&=&~ \frac{\partial E}{\partial t} + \kappa \, (\dot a B + \nabla a \times E) \label{effectiveCurrent} 
\end{eqnarray}
Note that if $a$ is a space-time constant then its contribution vanishes.  This might have been anticipated from the second form of Eqn.(\ref{axionEML}), since in that case we are adding a perfect derivative,
which does not contribute to the classical equations of motion.  

The new terms have simple physical interpretations.  According to Eqns.(\ref{effectiveCharge}, \ref{effectiveCurrent}) we have new contributions to the electric charge and current densities in the form
\begin{eqnarray}
\rho_a ~&=&~ - \kappa \, \nabla a \cdot B \nonumber \\
j_a ~&=&~ \kappa \,  (\dot a B + \nabla a \times E) 
\end{eqnarray}
In words: a magnetic field imparts a proportional electric dipole moment to the axion fields; the rest, including a corresponding displacement current and Hall effect, follows from special relativity.   

Now let us consider the possibility of a material with $a= a_0$ inside, interfacing with ``vacuum'', where $a=0$.  (The new term represents a kind of magneto-electric polarizability for our material.)  Let's suppose, for simplicity, that we have a sample of the material that has a constant profile in the $x-y$ plane, and terminates on flat faces at $z= 0, L$.  Then we will have $\delta$-function jumps in the gradients of Eqns.(\ref{effectiveCharge}, \ref{effectiveCurrent}).  Integrating across them, we see the possibility to induce surface charges and currents.   Thus for the top surface we have 
\begin{eqnarray}\label{axionResults}
\rho_a ~&=&~  a_0 \, \kappa  \,  B  ~=~   a_0 \,  \frac{e^2}{4\pi^2}  \,  B \nonumber \\
j_a ~&=&~ - a_0 \, \frac{e^2}{4\pi^2} \, \hat z \times E
\end{eqnarray}
and for the bottom surface, results of the opposite sign.   

We might also have worked directly from the action, integrating over $z$ to find the surface terms
\begin{eqnarray}
&{}& \int \, dt \, dx \, dy \, dz \ \epsilon^{\alpha \beta \gamma \delta} a(z) F_{\alpha \beta} F_{\gamma \delta} \nonumber \\ 
&{}& \propto \int \, dt \, dx \, dy \ \epsilon^{abc} A_a (t, x, y, z = L) \, \partial_b  A_c (t, x, y, z =L) \,  \nonumber \\
&{}& - \, \int \, dt \, dx \, dy \ \epsilon^{abc} A_a (t, x, y, z = 0) \, \partial_b  A_c (t, x, y, z =0)  
\end{eqnarray}
in an evident notation.   We get, in other words, two Chern-Simons theories, with opposite sign coefficients, located on the two boundaries.   

\subsubsection{Landau Levels Made Simple}

It will be revealing to compare these equations with the theory of Landau levels.  To emphasize the foundational nature of the relevant bits of that theory I'll derive it from first principles, in a few lines.

We are interested in electrons whose motion is confined to a two-dimensional $xy$ plane, under the influence of a strong orthogonal magnetic field of magnitude $B$.   Choosing our gauge so that $A_y =0$, and boldly throwing away the terms that do not grow with $B$, we have the particle Lagrangian 
\begin{eqnarray} 
L ~ &\approx& ~  \vec v \cdot \vec A  \nonumber \\
~&=&~ E \, B \, y \, \dot x
\end{eqnarray}
Thus $x$ is the only dynamical degree of freedom, and we have 
\begin{eqnarray}
p_x ~&=&~ e\, B \, y \nonumber \\
H ~&=&~ 0
\end{eqnarray}
We have a large degeneracy of states, in the form
\begin{equation}
N ~=~ L_x \int \frac{dp_x }{2\pi} ~=~ L_x L_y \frac{eB}{2\pi}
\end{equation}
which can support (if they're all occupied) charge density
\begin{equation}
\rho ~=~ eN/A ~=~  \frac{e^2 }{2\pi} \, B
\end{equation}

If we assume a fixed filling fraction $f$, we discover that application of a differential magnetic field $\Delta B (x,y)$ induces charge density according to
\begin{equation}\label{magneticResponse}
\Delta \rho (x, y) ~=~ f \ \frac{e^2 }{2\pi} \Delta B (x, y) 
\end{equation}

Now let us add in a planar electric field $E_y$ in the $\hat y$ direction.  This leads us to
\begin{eqnarray}
L ~&=&~ eB \, y \, \dot x \, - \, e E_y \, y \nonumber \\
H ~&=&~ eE_y \, y ~=~ \frac{E_y}{B} p_x \nonumber \\
\dot x ~&=&~  \frac{E_y}{B} 
\end{eqnarray}
and
\begin{equation}\label{electricResponse}
j_x~=~ \rho \dot x ~=~ f \frac{e^2 }{ 2 \pi} \, E_y
\end{equation}

\subsubsection{Interpretation and Applications}

Now we can bring our threads together.  The equations of axion electrodynamics, in Eqn.\,(\ref{axionResults}), match the Landau level responses Eqns.\,(\ref{magneticResponse}, \ref{electricResponse}), for
\begin{equation}
a_0 ~=~ 2 \pi f
\end{equation} 
-- with the significant difference that no background magnetic field is required (though there could be one).   In the quantum field theory of axions, we have periodicity under $a \rightarrow a + 2 \pi n$, for integer $n$.  This is the reason one often writes $a = \theta$ (or, in a theory with dynamical axions, $a = F \theta$, to achieve a normalized kinetic term).  Here we see that a shift in the jump $a_0$ across an interface by $2\pi$ corresponds to adding the response of a full Landau level of surface states.   Fractional values of $\frac{a_0}{2\pi}$ are more unusual.  They correspond, in the microscopic theory, to the response of a surface layer consisting of a wall of fractional charge, yielding a fractional Landau level.   

The time-reversal transformation $T$ takes $a \rightarrow -a$.  Taking into account the periodicity of $a$, this implies that there are two inequivalent possibilities consistent with $T$ symmetry, i.e. 
\begin{equation}
\frac{a}{2\pi} \equiv 0 \,  {\rm or\/} \, \frac{1}{2} \  {\rm (mod \ 1)}
\end{equation}
Topological insulators realize the non-trivial class.   Above, we saw how their effective theory is connected to a simple model of electron surface states, based on the theory of topological domain walls.

The effective theory of axion electrodynamics, both for topological insulators and more generally, can be used to describe the response of the material to real or virtual photons at low energy.  Thus it can be used to describe induced charges and Hall flows, as described above, and the forces they cause; a variety of optical effects, notably including Faraday and Kerr rotation \cite{faradayKerr} \cite{fKexpt}; and some especially interesting long range, Casimir-like forces.     On the other hand, many important physical effects lie beyond its scope, as does the theoretical determination of the value of $a_0$ for any particular material.

\section*{Acknowledgements}
This work is supported by the U.S. Department of Energy under grant Contract Number  DE-SC0012567.

\end{document}